\documentclass{article}

\PassOptionsToPackage{numbers,sort&compress}{natbib}
\usepackage[preprint]{neurips_2026}

\usepackage[utf8]{inputenc}
\usepackage[T1]{fontenc}
\usepackage{hyperref}
\usepackage{url}
\usepackage{booktabs}
\usepackage{amsmath}
\usepackage{amssymb}
\usepackage{amsfonts}
\usepackage{nicefrac}
\usepackage{microtype}
\usepackage{xcolor}
\usepackage{graphicx}
\usepackage{courier}
\usepackage{needspace}
\usepackage{wrapfig}
\usepackage{placeins}

\newcommand{\Dcal}{\mathcal{D}}

\title{Design-Ignoring versus Design-Respecting World Models for Epidemiology}

\author{%
  Xiangyu Yu \\
  Division of Biostatistics \\
  University of California, Berkeley\\
  Berkeley, CA 94720 \\
  \texttt{sean.yu@berkeley.edu} \\
\And
  Weiyu Liu \\
  School of Computer Science\\
  Georgia Institute of Technology\\
  Atlanta, GA 30332 \\
  \texttt{wliu490@gatech.edu} \\
}
\begin{document}
\maketitle

\begin{abstract}
World models for epidemiology learn from records shaped by study designs, including assignment, sampling, measurement, and related processes.  A model may therefore reconstruct observed trajectories while learning an intervention contrast that depends on how records were collected.  We formalize study design as constraints on a world model's latent-world interface, action mechanism, observation likelihood, and target readout.  This distinguishes \emph{design-respecting} models, which encode these constraints, from \emph{design-ignoring} models, which fit selected records without relating assignment and observation to the intended intervention question.  Using a large scale cluster-randomized test-negative trial, we hold the latent structure and fitting settings fixed and compare a design-ignoring case-count model with a design-respecting test-negative observation model.  Both models achieve comparable factual reconstruction.  Yet across 500 paired resampling experiments at 80\%, 60\%, and 40\% relative sampling intensity, median contrast changes are $-19.96\%$, $-41.06\%$, and $-60.98\%$ for the design-ignoring model, versus $-0.95\%$, $-3.00\%$, and $-5.32\%$ for the design-respecting model.  The same qualitative separation holds when sampling also varies across clusters.  Thus, factual reconstruction alone does not establish design alignment; testing whether fitted contrasts preserve design-implied observation-process invariances provides a sharper evaluation.
\end{abstract}

\section{Introduction}

World models represent latent state and action-conditioned dynamics for prediction, planning, and imagined rollouts \citep{ha2018world,hafner2019planet,hafner2025dreamerv3,ding2024survey}.  In biomedicine, this becomes intervention-conditioned simulation of molecular, physiological, or patient states \citep{chen2026medical,yang2025mewm,mu2026ehrworld}.  The distinction from traditional statistics is functional rather than algorithmic: classical state-space, mechanistic, and agent-based models can also support such simulations.

Memon et al.\ separate latent epidemic state from policy-dependent surveillance, including testing, delays, noise, and behavioral response \citep{memon2026}.  Surveillance, however, is not the whole data-generating process: study design determines entry, assignment, testing, retained observations, and the contrast it identifies.  Related work motivates causal grounding for intervention queries \citep{chen2026medical,hernan2016target,noorbakhsh2022counterfactual}.  We focus on a complementary issue: making study design a first-class world-model component.

Our contributions are a component-level framework for translating study design into world-model construction and a paired evaluation strategy that tests a study-design-implied observation-process invariance.  Using participant data from the Applying Wolbachia to Eliminate Dengue (AWED) trial \citep{anders2018protocol,utarini2021}, we derive a design-respecting emission while retaining the state specification.  The results demonstrate why factual reconstruction is insufficient to justify an intervention interpretation.  Study design is therefore central to what a world model's outputs mean and how they should be evaluated.

\section{Design-ignoring and design-respecting world models}
\label{sec:alignment}

\subsection{Conventional world-model formulation}

Let $x_t$ denote the latent state, $a_t$ the action, and $o_t$ the observation.
Let $\mathcal H_t=(o_{1:t},a_{1:t-1})$ denote raw history and
$h_t=f_\phi(\mathcal H_t)$ a recurrent or belief-state representation.  A controlled model is
\begin{equation}
  x_1\sim p_\phi,\quad x_{t+1}\sim P_\phi(\cdot\mid x_t,a_t),
  \quad o_t\sim\Omega_\phi(\cdot\mid x_t,a_{t-1}).
  \label{eq:conventionalwm}
\end{equation}
Such models learn transitions and emissions from trajectories and simulate
under actions \citep{kaelbling1998,ha2018world,hafner2019planet,krishnan2017structured}.
We use ``design-ignoring'' for models that fit selected records without
relating assignment and observation to the target.  A
clinic-count emission describes recorded cases; population incidence also
depends on who was observed.

\subsection{Mapping study design to world-model components}

Following standard epidemiologic accounts of study design \citep{lash2021modern}, we summarize the components relevant to world-model construction as
\begin{equation}
  \Dcal=(\mathcal U,G,E,S,M,\Psi,\ldots),
  \label{eq:design}
\end{equation}
where $\mathcal U$ defines study units, eligibility, source and target populations, and time structure; $G$ the treatment-assignment or exposure-adoption mechanism; $E$ the mapping from assignment or adoption to realized exposure, including adherence and interference; $S$ sampling, selection, censoring, and retention; $M$ measurement of exposures, covariates, and outcomes; $\Psi$ the estimand or prediction target; and $\ldots$ any additional components required by a particular study design.  Together, these study-design specifications constrain the components already present in Eq.~\eqref{eq:conventionalwm}:
\begin{equation}
 \begin{split}
 h_t&=f_\phi(\mathcal H_t),\quad x_1\sim p_\phi,\quad
 a_t\sim\pi_{\Dcal}(\cdot\mid h_t),\\
 x_{t+1}&\sim P_\phi(\cdot\mid x_t,a_t),\qquad
 o_t\sim\Omega_{\phi,\Dcal}(\cdot\mid x_t,a_{t-1}).
 \end{split}
 \label{eq:designwm}
\end{equation}
\paragraph{Specify the latent-world interface.}
$\mathcal U$ and $E$ define the meaning and resolution of $x_t$, $a_t$, and $\mathcal H_t$ before fitting, including whether action means assignment, realized treatment, or mapped cluster/network exposure.  The encoder $f_\phi$, initial-state law $p_\phi$, and transition $P_\phi$ are then specified relative to this interface.

\paragraph{Align the action mechanism.}
When actions are generated rather than externally set, $G$ constrains $\pi_{\Dcal}(a_t\mid h_t)$.  Trials supply known randomization; observational treatment may follow measured history; and quasi-experimental adoption may follow a cutoff, instrument, or calendar rule.  Intervention rollouts set $a_t$ according to the specified intervention rather than sample it from the observed policy \citep{hernan2016target}.

\paragraph{Observation, readout, and evaluation.}
$S$ and $M$ define $o_t$ and constrain $\Omega_{\phi,\Dcal}$ through conditioning, known weights, measurement error, or nuisance-factor cancellation; this observation law may operate on single times or whole trajectories.  $\Psi$ fixes the readout, target-population averaging, uncertainty unit, and checks compatible with the study design.  Observation changes should be separated from action changes, and assumption violations tested rather than reinterpreted as disease dynamics.  These constraints on $\pi_{\Dcal}$, $\Omega_{\phi,\Dcal}$, and the readout are the operational distinction between design-ignoring and design-respecting models: they determine the likelihood used for learning and the interpretation tested at evaluation.

\paragraph{What may enter $\Dcal$.}
Experiments may specify individual/cluster randomization, crossover washout, stepped-wedge timing, adaptive allocation, adherence, and interference.  Observational studies may specify cohort eligibility and censoring, exposure/confounder histories, or outcome-dependent case-control or test-negative sampling.  Quasi-experiments may specify a regression-discontinuity cutoff, an instrument and exclusion restriction, or intervention timing, comparison series, and trend assumptions.  Study designs are often mixed: AWED combines cluster randomization with test-negative sampling \citep{anders2018crtnd}; pragmatic trials may use routine-care measurements, and natural experiments may add survey sampling.  A label such as ``trial'' therefore does not specify $\Dcal$.

\section{Case study: a cluster-randomized test-negative design}
\label{sec:awed}

AWED randomized 24 Yogyakarta clusters to Wolbachia deployment or control and recruited eligible febrile patients through a test-negative design \citep{anders2018protocol,utarini2021}.  The data contain 6,306 deduplicated participant-events over 27 months. We distinguish cluster $c$, the assignment and resampling unit, and month $t$, the dynamic time unit.  Let $Z_c\in\{0,1\}$ denote residential cluster assignment, $Y^\pm_{ct}$ the enrolled positive and negative counts, and $N_{ct}=Y^+_{ct}+Y^-_{ct}$.  Both models are fitted to all $24\times27$ cluster-month cells, including zeros. We use a deliberately simple state-space model to isolate the consequences of design alignment from architectural complexity: we first model the recorded dengue case trajectories, then incorporate the study design into the observation model.

\subsection{Design-ignoring case-count model}
\label{sec:ignoring}

Take $o_{ct}=Y^+_{ct}$ and the action input to be the time-constant assignment $Z_c$.  A shared monthly state $x_t$ describes temporal variation, while a cluster intercept $u_c$ captures persistent heterogeneity:
\begin{equation}
\begin{split}
  x_1&\sim N(0,\sigma_x^2),\quad x_{t+1}\mid x_t\sim N(\rho x_t,\sigma_x^2),\quad
  u_c\sim N(0,\sigma_u^2),\\
  Y^+_{ct}\mid x_t,Z_c,u_c&\sim\operatorname{Poisson}(\mu_{ct}),\qquad
  \log\mu_{ct}=\alpha_0+\alpha_Z Z_c+x_t+u_c.
\end{split}
\label{eq:awedignoringwm}
\end{equation}
This specifies the initial state, transition, and emission of Eq.~\eqref{eq:conventionalwm}, with assignment affecting the emission.  The positive-count mean combines disease occurrence and ascertainment; negative events do not contribute to its likelihood.  Factual fit compares $\widehat\mu_{ct}$ with observed $Y^+_{ct}$, and $\exp(\widehat\alpha_Z)$ is the conditional count ratio at a common state and cluster intercept.  Propagating the state and sampling the emission generates positive-count trajectories at a specified assignment.

\subsection{Design-respecting TND observation model}
\label{sec:respecting}

AWED's one-time constrained randomization fixes the action $Z_c$, and fitting conditions on the realized allocation \citep{anders2018protocol}.  Conditioning on enrollment, we use the cluster-month observation model
\begin{equation}
 Y^+_{ct}\mid N_{ct},x_t,Z_c,u_c\sim\operatorname{Binomial}(N_{ct},p_{ct}),\quad \operatorname{logit}p_{ct}=\beta_0+\beta_Z Z_c+x_t+u_c.
 \label{eq:binomial}
\end{equation}
Negative events therefore enter through the enrollment total rather than being discarded.  The readout is the conditional odds ratio $\exp(\beta_Z)$, with fitted positive counts $N_{ct}\widehat p_{ct}$; the models retain the same latent/random-intercept structure and differ only in the observation likelihood.

\subsection{Testing the TND-implied observation-process invariance}
\label{sec:sampling}

The TND uses test-negative controls to separate disease effects from variation in health-care seeking and ascertainment: under its identifying assumptions, shared relative ascertainment cancels from the positive-versus-negative odds contrast \citep{anders2018crtnd,dufault2020,jewell2019}.  We therefore test design alignment by asking whether each world model inherits this invariance when observation intensity changes equally for positive and negative events.

For each of 500 arm-stratified cluster-bootstrap replicates, we fit both models to a reference sample and then perturb only intervention-arm observation at relative intensity $r\in\{0.8,0.6,0.4\}$, retaining positive and negative events with the same cluster-month probability.  Sampling varies over time in the monthly scenario and additionally across clusters in the cluster-month scenario; Appendix~\ref{app:sampling} gives the full calibrated policies.  This creates arm-differential ascertainment while preserving the TND condition that relative ascertainment is shared across positive and negative events; assignment, eligibility, laboratory classification, and the study window remain unchanged.

Let $\widehat C^{(\mathrm{ign})}=\exp(\widehat\alpha_Z)$ and $\widehat C^{(\mathrm{resp})}=\exp(\widehat\beta_Z)$.
\begin{equation}
 \Delta_b^{(m)}(r)=\widehat C_{b,\mathrm{sample}}^{(m)}(r) / \widehat C_{b,\mathrm{ref}}^{(m)}-1.
 \label{eq:pairedchange}
\end{equation}
We report $\Delta$ as a percentage.  Because the models target different conditional contrasts, this paired normalization compares their sensitivity to observation changes rather than their contrast values directly.

\Needspace{8\baselineskip}
\subsection{Factual fit and invariance results}

\begin{wrapfigure}[18]{r}{0.45\textwidth}
\vspace{-4em}
 \centering
 \includegraphics[width=\linewidth]{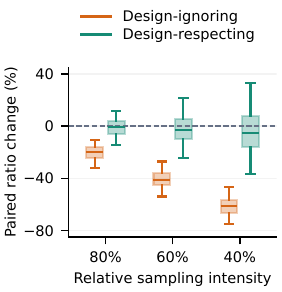}
 \caption{Shared monthly sampling. Boxes: central 50\% and median line; whiskers: central 95\% of 500 replicates.}
 \label{fig:shift}
\end{wrapfigure}

The factual conditional count and odds ratios are 0.2275 and 0.2492, respectively, both numerically close to the published AWED primary ITT aggregate TND odds ratio of 0.23 (95\% CI 0.15--0.35) \citep{utarini2021}.  Because the published value is an aggregate TND odds ratio whereas ours are model-based conditional contrasts, we use this comparison only as a numerical sanity check.  Cluster-month positive-count RMSE is 0.940 and 0.882, respectively; Appendix~\ref{app:checks} gives bootstrap intervals and supporting checks.

Under shared monthly sampling, percentage median changes (central 95\% ranges) for design-ignoring versus design-respecting are $-19.96$ $(-32.52,-10.24)$ versus $-0.95$ $(-14.54,12.05)$ at 80\% intensity, $-41.06$ $(-54.00,-27.00)$ versus $-3.00$ $(-24.39,21.87)$ at 60\% intensity, and $-60.98$ $(-74.87,-46.23)$ versus $-5.32\%$ $(-36.68,33.46)$ at 40\% intensity (Figure~\ref{fig:shift}).

Adding cluster-specific sampling variation gives the same qualitative separation; at 60\% intensity the medians are $-41.72\%$ and $-2.88\%$, respectively (Appendix~\ref{app:heterogeneity}).

Thus, the design-respecting contrast substantially better preserves the TND-implied observation-process invariance despite comparable factual reconstruction.

\section{Discussion}

Study design supplies concrete instructions for world-model construction.  In AWED, cluster randomization fixes the action mechanism, while recruitment and testing motivate a likelihood conditional on enrollment.  Holding the cluster structure and dynamics fixed deliberately isolates this observation component and shows how it changes fitted contrasts under shared sampling; in other settings, realized exposure may also alter the latent transition $P_\phi$.  TND supplies the established statistical structure \citep{anders2018crtnd,dufault2020}; our framework connects it to world-model components and design-grounded evaluation.  More broadly, factual fit should be accompanied by design-grounded checks that perturb sampling or measurement while holding the intervention fixed.  Specifying observation mechanisms, target estimands, and failure conditions helps distinguish surveillance artifacts from intervention effects and clarifies what model outputs can support.

\section{Limitations, impacts, and responsible use}

The model uses a static cluster random intercept and a shared monthly trajectory; spatial interference and cluster-specific epidemic dynamics require extensions \citep{johnson2025spatiotemporal}.  Its conditional odds ratios have a disease interpretation only under the stated TND assumptions and are not population-averaged effects \citep{jewell2019}.  The case study conditions on observed enrollment rather than modeling its dynamics, so it neither identifies absolute incidence nor generates clinic volume without additional denominator and observation-process information.  The illustration covers one cluster-randomized TND; other study designs require separate instantiation and validation.  The sampling policies are imposed on observed records, and finite-sample refitting produces residual uncertainty even under shared sampling; bootstrap inference is further limited by the unavailable constrained-allocation set.

\appendix

\section{Reproducibility and experimental details}
\label{app:repro}

\paragraph{Data and statistical units.}
We remove duplicate serotype rows by participant identifier, keeping one binary test result per participant-event, and derive calendar month from illness onset.  The resulting data contain 6,306 events (385 positive and 5,921 negative), 24 residential clusters, and 27 months.  We construct all 648 cluster-month cells, including zero-event months.  Models are fitted at this level; aggregation across clusters is used only to display factual trajectories.  The 12 intervention and 12 control clusters are the units of bootstrap resampling.  All observed months are used for factual reconstruction and the paired sampling experiment.

\paragraph{Matched latent structure and estimation.}
Both models use independent $u_c\sim N(0,\sigma_u^2)$ and
\begin{equation}
 x_1\sim N(0,\sigma_x^2),\qquad x_t\mid x_{t-1}\sim N(\rho x_{t-1},\sigma_x^2),
 \label{eq:state}
\end{equation}
with $\rho=0.9$, $\sigma_x^{-2}=5$, and fixed-effect ridge $\kappa=10^{-4}$.  These settings are common to both models; no hyperparameter search is performed.  For model $m$, write $\gamma=(\gamma_0,\gamma_Z)$ for its fixed effects and $f_m$ for its conditional emission probability mass function.  We minimize
\begin{equation}
\begin{split}
 \mathcal J_m(\gamma,x,\sigma_u)={}&-\sum_c\log\int_{\mathbb R}
 \left\{\prod_t f_m(Y^+_{ct}\mid N_{ct},x_t,Z_c,u)\right\}
 \varphi(u;0,\sigma_u^2)\,du\\
 &+\frac{1}{2\sigma_x^2}\left[x_1^2+\sum_{t=2}^T(x_t-\rho x_{t-1})^2\right]
 +\frac{\kappa}{2}(\gamma_0^2+\gamma_Z^2),
\end{split}
\label{eq:fitobjective}
\end{equation}
where $f_{\rm ign}$ is Poisson and does not use $N_{ct}$, while $f_{\rm resp}$ is binomial conditional on $N_{ct}$.  The models have two fixed effects, 27 fitted monthly states, and one estimated cluster variance.  The 24 cluster intercepts are integrated out, and their conditional modes are used for factual fitted counts.  Each model estimates its own parameters and states.

The one-dimensional integrals are approximated by Laplace's method at each cluster's conditional mode.  Newton iteration computes these modes with step tolerance $10^{-9}$ and a maximum of 100 iterations.  An analytic gradient includes the derivative of the Laplace correction.  L-BFGS-B optimizes the resulting objective with relative objective tolerance $10^{-11}$, gradient tolerance $2\times10^{-6}$, and at most 1,500 iterations.  Fits with an unsuccessful termination or fixed-effect/state gradient above 0.003 are restarted at tolerance $10^{-13}$ and at most 2,500 iterations.  We estimate $\sigma_u$ anew in every fit, with numerical bounds $\log\sigma_u\in[-7,2]$.  All reported fits converged, and no cluster-variance lower boundary was reached.  Appendix~\ref{app:numerical} checks integration accuracy.

\subsection{Relative sampling-intensity perturbation}
\label{app:sampling}

For replicate $b$, we sample 12 clusters with replacement within each arm, keeping all 27 months.  Repeated selections receive distinct bootstrap cluster identities and separate random intercepts in fitting.  Let $Y^\pm_{ct,b}$ and $N_{ct,b}$ denote the counts in this reference sample.  We first fit both models to obtain their respective reference coefficients.  The same reference fits provide the 500-replicate percentile intervals for factual contrasts.

For the monthly policy, generate a logit-scale path $g_{tb}$:
\begin{equation}
\begin{split}
 g_{1b}&\sim N(0,0.8^2),\qquad
 g_{tb}=0.7g_{t-1,b}+\epsilon_{tb},\\
 \epsilon_{tb}&\sim N\!\left(0,0.8^2(1-0.7^2)\right),\qquad
 s_{tb}(r)=\operatorname{logit}^{-1}(a_{rb}+g_{tb}).
\end{split}
\label{eq:samplingpolicy}
\end{equation}
For the cluster-month policy, additionally draw independent cluster offsets $v_{cb}\sim N(0,0.8^2)$ and set $s_{ctb}(r)=\operatorname{logit}^{-1}(a_{rb}+g_{tb}+v_{cb})$.  These policy offsets are distinct from the model intercepts $u_c$.  Copies of a source cluster share its sampling offset, while event sampling is independent for each copy.  We use the same paths and offsets across intensities, and calibrate a separate intercept for each policy and $r\in\{0.8,0.6,0.4\}$ so that
\begin{equation}
 \frac{\sum_{c:Z_c=1}\sum_t N_{ct,b}s_{ctb}(r)}
      {\sum_{c:Z_c=1}\sum_t N_{ct,b}}=r.
 \label{eq:samplingcalibration}
\end{equation}
For the monthly policy, $s_{ctb}(r)=s_{tb}(r)$.  Thus $r$ is the reference-event-weighted expected sampled fraction, termed relative sampling intensity; realized fractions vary across replicates.

For each intervention cluster-month, independently draw
\begin{equation}
 \widetilde Y^\pm_{ct,b}\mid Y^\pm_{ct,b}
 \sim\operatorname{Binomial}\!\left(Y^\pm_{ct,b},s_{ctb}(r)\right).
 \label{eq:thinning}
\end{equation}
This implements Bernoulli event sampling with a common probability for both labels.  Control counts remain unchanged.  We refit both models to $(\widetilde Y^+_{ct,b},\widetilde Y^-_{ct,b})$ at the cluster-month level and compute Eq.~\eqref{eq:pairedchange}.  The 500 replicates capture whole-cluster resampling, random policy variation, and event sampling.  Boxes and whiskers in Figure~\ref{fig:shift} summarize these sensitivity distributions.  Sampling operates within the original study window; we do not redefine the window using the positive events remaining after sampling.

The main seed sequence is \texttt{[20260910, 0]}.  The policy parameters are specified in advance.  Implementation checks verify count validity, sampling-intensity calibration, optimization convergence, and agreement with the corresponding shared-sampling calculation.

\paragraph{Software and compute.}
The two-scenario run uses Python 3.13.3, NumPy 2.5.2, pandas 3.0.5, SciPy 1.18.1, and Matplotlib 3.11.1 on the local Apple M1 computer (8 CPU cores, 16 GB memory), with four CPU worker processes.  It performs 7,002 main fits: two factual fits, 1,000 reference fits, and 6,000 sampled fits over two policies, three intensities, two models, and 500 replicates.  The recorded fitting time is 14.5 seconds, excluding figure rendering and the additional numerical checks.  Earlier exploratory cluster-month runs and validation were also CPU-only and required less than ten additional CPU-minutes.  No GPU or cloud compute is used.

\section{Supporting analyses and validation}
\label{app:checks}

\subsection{Factual reconstruction}

\begin{figure}[ht]
 \centering
 \includegraphics[width=\linewidth]{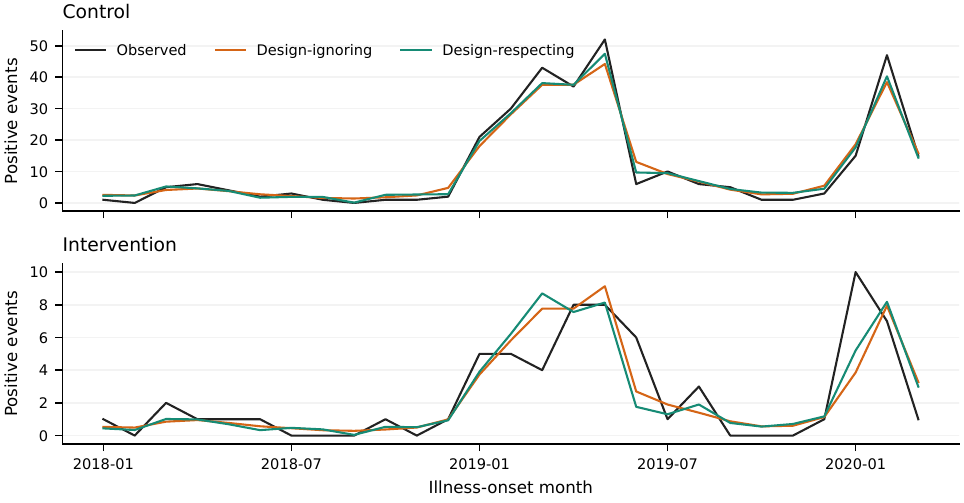}
 \caption{Observed and fitted positive events, summed across clusters for display. Predictions are obtained from cluster-month fits; the corresponding arm-month display RMSE values are 2.581 and 2.035.}
 \label{fig:factual}
\end{figure}

Table~\ref{tab:factual} reports the two conditional contrasts, percentile intervals from 500 arm-stratified whole-cluster bootstrap replicates, and positive-count reconstruction error.  Models are fitted at the cluster-month level throughout.  Figure~\ref{fig:factual} aggregates predictions across clusters only for visualization.

\begin{table}[ht]
 \centering\small
 \caption{Factual conditional contrasts and reconstruction. RMSE is measured in positive events per cluster-month; intervals are whole-cluster bootstrap percentiles.}
 \label{tab:factual}
 \begin{tabular}{lrrrr}
 \toprule
 Model & Conditional ratio & 95\% interval & $\widehat\sigma_u$ & RMSE\\
 \midrule
 Design-ignoring & 0.2275 & [0.1464, 0.3517] & 0.4876 & 0.9400\\
 Design-respecting & 0.2492 & [0.1681, 0.3848] & 0.4001 & 0.8815\\
 \bottomrule
 \end{tabular}
\end{table}

\FloatBarrier

\subsection{TND identification conditions and sampling invariance}
\label{app:tnd}

\citet{anders2018crtnd} distinguish six conditions: the intervention does not affect negative-control illness; relative health-care seeking or enrollment across intervention groups is shared by the two illness categories; intervention efficacy is not modified by care-seeking behavior when generalizing beyond clinic attendees; diagnostic classification is sufficiently sensitive and specific; control sampling is inclusive of people who may be cases at another time; and controls are sampled during periods when the target illness is circulating.  The two experimental policies apply the same sampling probability to both labels within every cluster-month, while preserving the original assignment, calendar time, eligibility rules, and recorded classification.  They impose no disease intervention.  Consequently, they preserve the shared-relative-ascertainment condition at the cell level if it holds in the source process; they do not independently establish all source-study assumptions.

To state the conditional interpretation, let $\lambda^\pm_{ct}(z)$ denote the underlying illness intensities under assignment $z$ and $q^\pm_{ct}(z)$ the corresponding ascertainment probabilities.  Within a fixed cluster-time context, the negative-control and relative-ascertainment conditions are
\begin{equation}
 \lambda^-_{ct}(1)=\lambda^-_{ct}(0),\qquad
 \frac{q^+_{ct}(1)}{q^+_{ct}(0)}=\frac{q^-_{ct}(1)}{q^-_{ct}(0)}.
 \label{eq:tndassumptions}
\end{equation}
For observed intensities $\nu^\pm_{ct}(z)=q^\pm_{ct}(z)\lambda^\pm_{ct}(z)$, these give
\begin{equation}
 \frac{\nu^+_{ct}(1)/\nu^-_{ct}(1)}{\nu^+_{ct}(0)/\nu^-_{ct}(0)}
 =\frac{\lambda^+_{ct}(1)}{\lambda^+_{ct}(0)}.
 \label{eq:tndratio}
\end{equation}
The fitted logistic coefficient specifies a common conditional odds ratio, $\exp(\beta_Z)$.  Under the corresponding conditional TND interpretation, Eq.~\eqref{eq:tndratio} links this to a relative illness intensity.  A population-averaged effect is a separate target requiring its own averaging and interpretation \citep{jewell2019}.  Randomization supplies the assignment mechanism; the sampling experiment assesses each fitted contrast relative to its reference.

For any given cluster-month, shared event sampling multiplies both observed intensities by $s_{ct}$ and leaves their conditional positive probability unchanged:
\begin{equation}
 \frac{s_{ct}\nu^+_{ct}}{s_{ct}\nu^+_{ct}+s_{ct}\nu^-_{ct}}
 =\frac{\nu^+_{ct}}{\nu^+_{ct}+\nu^-_{ct}}.
 \label{eq:samplingcancellation}
\end{equation}
This is the observation-process property motivated by TND \citep{dufault2020}.  Conditional independence of event labels with common $p_{ct}$ within a cell is the working likelihood assumption; normal random intercepts and AR(1) states are additional modeling choices.  Sampling reduces information and changes the weights of cells, so refitted coefficients can move through finite-sample variation, variance estimation, and penalization.  Keeping clusters separate avoids imposing a common positive probability across all clusters in an arm-month, an issue relevant to heterogeneous ascertainment \citep{wang2023}.

\subsection{Results across sampling policies and intensities}
\label{app:heterogeneity}

\begin{figure}[ht]
 \centering
 \includegraphics[width=0.64\linewidth]{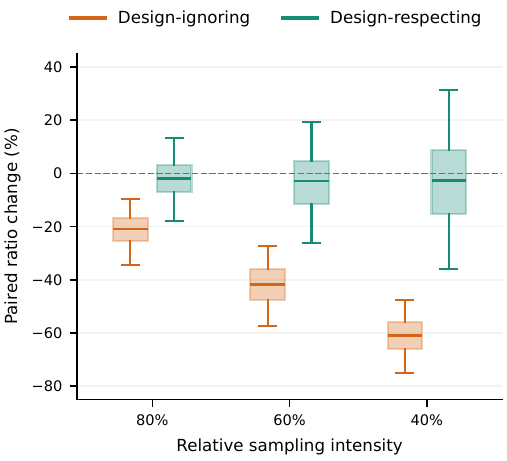}
 \caption{Shared cluster-month sampling. Adding cluster-specific sampling variation yields the same qualitative contrast as shared monthly sampling in Figure~\ref{fig:shift}. Boxes show medians and the middle 50\%; whiskers span the central 95\% of 500 paired replicates.}
 \label{fig:cluster_sampling}
\end{figure}

Table~\ref{tab:sampling} reports both shared-sampling policies at all three intensities.  At 60\% intensity under cluster-month sampling, the median conditional-odds-ratio change is $-2.88\%$ (central 95\% range $[-26.29\%,19.54\%]$), compared with $-41.72\%$ for the count ratio.  The same qualitative separation holds across all three intensities (Figure~\ref{fig:cluster_sampling}).

\begin{table}[ht]
\centering\small
\caption{Median paired ratio changes (percent), with central 95\% ranges over 500 replicates. Both labels share a sampling probability within each cluster-month.}
\label{tab:sampling}
\begin{tabular}{llrr}
\toprule
Policy & Intensity & Design-ignoring & Design-respecting\\
\midrule
Monthly & 80\% & $-19.96$ [$ -32.52, -10.24 $] & $-0.95$ [$ -14.54, 12.05 $]\\
Monthly & 60\% & $-41.06$ [$ -54.00, -27.00 $] & $-3.00$ [$ -24.39, 21.87 $]\\
Monthly & 40\% & $-60.98$ [$ -74.87, -46.23 $] & $-5.32$ [$ -36.68, 33.46 $]\\
Cluster-month & 80\% & $-20.98$ [$ -34.50, -9.56 $] & $-1.87$ [$ -17.88, 13.36 $]\\
Cluster-month & 60\% & $-41.72$ [$ -57.38, -27.37 $] & $-2.88$ [$ -26.29, 19.54 $]\\
Cluster-month & 40\% & $-60.98$ [$ -75.12, -47.68 $] & $-2.70$ [$ -36.09, 31.51 $]\\
\bottomrule
\end{tabular}
\end{table}

\subsection{Numerical validation}
\label{app:numerical}

Central finite differences verify the analytic gradients, with maximum absolute errors $4.0\times10^{-9}$ and $3.4\times10^{-8}$ for the Poisson and binomial models.  Replacing the Laplace approximation with 25-node adaptive Gauss--Hermite quadrature gives conditional ratios 0.22758 and 0.24935, compared with 0.22747 and 0.24924 under Laplace.  For the preselected bootstrap replicate 0 at 40\% intensity under both policies, relative ratio differences remain below 0.074\% and paired-change differences below 0.025 percentage points.  All 6,000 sampled fits pass convergence checks, with maximum fixed-effect/state gradient 0.0026.

\section{Data access, licensing, and ethics}
\label{app:assets}

\subsection{Data and code availability}
\label{app:data-code}

Participant-level AWED records are not redistributed under their original access conditions.  The manuscript specifies preprocessing, model equations, numerical settings, perturbations, random seeds, and software versions needed to reconstruct the analysis.  Following acceptance, analysis code and reproduction instructions will be released under the MIT License.

\subsection{Existing assets and licensing}
\label{app:licensing}

The AWED protocol is published under CC BY 4.0, with the article's associated data covered by CC0 unless otherwise stated \citep{anders2018protocol}.  The AWED \emph{New England Journal of Medicine} author manuscript is available under a CC BY license \citep{utarini2021}, and the subsequent AWED spatiotemporal analysis used for scientific context is published under CC BY 4.0 \citep{johnson2025spatiotemporal}.  These works are attributed and cited; third-party participant-level records are neither redistributed nor relicensed.

\subsection{Ethics and participant protection}
\label{app:ethics}

This work is a derivative secondary analysis of de-identified records from the previously conducted and published AWED study, with no new recruitment, intervention, participant contact, or linkage to direct identifiers.  The original study obtained community approval, written informed consent from participants or guardians and assent from participants aged 13--17 years, and approval from the human research ethics committees at Universitas Gadjah Mada and Monash University \citep{anders2018protocol,utarini2021}.  This methodological reanalysis uses the established study data product and creates no additional physical or intervention risk.  Its residual participant risk is confidentiality loss, mitigated through de-identification, aggregate reporting, and non-release of participant-level records.

\end{document}